\documentclass{article}

\usepackage{microtype}
\usepackage{graphicx}
\usepackage{booktabs} 

\usepackage{hyperref}
\usepackage{balance}

\usepackage[accepted]{icml2018}

\usepackage{comment}
\usepackage{enumitem}
\usepackage{amsmath,amsfonts,mathtools}
\usepackage{subcaption}
\usepackage{tikz,tkz-graph}
\usetikzlibrary{
  graphs,
  graphs.standard,
  positioning,chains,fit,shapes,calc
}
\definecolor{nice-red}{HTML}{E41A1C}
\definecolor{nice-orange}{HTML}{FF7F00}
\definecolor{nice-yellow}{HTML}{FFC020}
\definecolor{nice-green}{HTML}{4DAF4A}
\definecolor{nice-blue}{HTML}{377EB8}
\definecolor{nice-purple}{HTML}{984EA3}

\usepackage{etoolbox}
\newcommand{\zerodisplayskips}{%
  \setlength{\abovedisplayskip}{1pt}%
  \setlength{\belowdisplayskip}{1pt}%
  \setlength{\abovedisplayshortskip}{1pt}%
  \setlength{\belowdisplayshortskip}{1pt}}
\appto{\normalsize}{\zerodisplayskips}
\appto{\small}{\zerodisplayskips}
\appto{\footnotesize}{\zerodisplayskips}

\makeatletter
\newcommand{\subalign}[1]{%
  \vcenter{%
    \Let@ \restore@math@cr \default@tag
    \baselineskip\fontdimen10 \scriptfont\tw@
    \advance\baselineskip\fontdimen12 \scriptfont\tw@
    \lineskip\thr@@\fontdimen8 \scriptfont\thr@@
    \lineskiplimit\lineskip
    \ialign{\hfil$\m@th\scriptstyle##$&$\m@th\scriptstyle{}##$\crcr
      #1\crcr
    }%
  }
}
\makeatother

\usepackage[accepted]{icml2018}

\icmltitlerunning{Learning Policy Representations in Multiagent Systems}

\begin{document}
\twocolumn[
\icmltitle{Learning Policy Representations in Multiagent Systems}



\icmlsetsymbol{equal}{*}

\begin{icmlauthorlist}
\icmlauthor{Aditya Grover}{stan}
\icmlauthor{Maruan Al-Shedivat}{cmu}
\icmlauthor{Jayesh K. Gupta}{stan}
\icmlauthor{Yura Burda}{open}
\icmlauthor{Harrison Edwards}{open}
\end{icmlauthorlist}

\icmlaffiliation{stan}{Stanford University}
\icmlaffiliation{cmu}{Carnegie Mellon University}
\icmlaffiliation{open}{OpenAI}

\icmlcorrespondingauthor{Aditya Grover}{adityag@cs.stanford.edu}

\icmlkeywords{multiagent systems, agent modeling, representation learning}

\vskip 0.3in
]



\printAffiliationsAndNotice{}  

\begin{abstract}  
Modeling agent behavior is central to understanding the emergence of complex phenomena in multiagent systems.
Prior work in agent modeling has largely been task-specific and driven by hand-engineering domain-specific prior knowledge.
We propose a general learning framework for modeling agent behavior in any multiagent system using only a handful of interaction data.
Our framework casts agent modeling as a representation learning problem.
Consequently, we construct a novel objective inspired by imitation learning and agent identification and design an algorithm for unsupervised learning of representations of agent policies.
We demonstrate empirically the utility of the proposed framework in (i) a challenging high-dimensional competitive environment for continuous control and (ii) a cooperative environment for communication, on supervised predictive tasks, unsupervised clustering, and policy optimization using deep reinforcement learning.
\end{abstract}

\section{Introduction}

Intelligent agents rarely act in isolation in the real world and often seek to achieve their goals through interaction with other agents.
Such interactions give rise to rich, complex behaviors formalized as per-agent policies in a multiagent system~\citep{ferber1999multi,wooldridge2009introduction}.
Depending on the underlying motivations of the agents, interactions could be directed towards achieving a shared goal in a collaborative setting, opposing another agent in a competitive setting, or be a mixture of these in a setting where agents collaborate in teams to compete against other teams.
Learning useful representations of the policies of agents based on their interactions is an important step towards characterization of the agent behavior and more generally inference and reasoning in multiagent systems.

In this work, we propose an unsupervised encoder-decoder framework for learning continuous representations of agent policies given access to only a few episodes of interaction.
For any given agent, the representation function is an encoder that learns a mapping from an interaction (\textit{i.e.}, one or more episodes of observation and action pairs involving the agent) to a continuous embedding vector.
Using such embeddings, we condition a policy network (decoder) and train it simultaneously with the encoder to \textit{imitate} other interactions involving the same (or a coupled) agent. Additionally, we can explicitly \textit{discriminate} between the embeddings corresponding to different agents using triplet losses.

For the embeddings to be useful, the representation function should \textit{generalize} to both unseen interactions and unseen agents for novel downstream tasks.
Generalization is well-understood in the context of supervised learning where a good model is expected to attain similar train and test performance.
For multiagent systems, we consider a notion of generalization based on \textit{agent-interaction graphs}.
An agent-interaction graph provides an abstraction for distinguishing the agents (nodes) and interactions (edges) observed during training, validation, and testing.

Our framework is agnostic to the nature of interactions in multiagent systems, and hence broadly applicable to competitive and cooperative environments. In particular, we consider two multiagent environments: (i) a competitive continuous control environment, 
\texttt{RoboSumo}~\citep{al2017continuous},
and (ii) a \texttt{ParticleWorld} environment of cooperative communication where agents collaborate to achieve a common goal~\citep{mordatch2017emergence}.
For evaluation, we show how representations learned by our framework are effective for downstream tasks that include clustering of agent policies (unsupervised), classification such as win or loss outcomes in competitive systems (supervised), and policy optimization (reinforcement).
In the case of policy optimization, we show how these representations can serve as privileged information for better training of agent policies.
In \texttt{RoboSumo}, we train agent policies that can condition on the opponent's representation and achieve superior win rates much more quickly as compared to an equally expressive baseline policy with the same number of parameters.
In \texttt{ParticleWorld}, we train speakers that can communicate more effectively with a much wider range of listeners given knowledge of their representations.

\section{Preliminaries}

In this section, we present the necessary background and notation relevant to the problem setting of this work.

\textbf{Markov games.}
We use the classical framework of Markov games \citep{littman1994markov} to represent multiagent systems.
A \textit{Markov game} extends the general formulation of partially observable Markov decision processes (POMDP) to the multiagent setting.
In a Markov game, we are given a set of $n$ agents on a state-space $\mathcal{S}$ with action spaces $\mathcal{A}_1, \mathcal{A}_2, \cdots, \mathcal{A}_n$ and observation spaces $\mathcal{O}_1, \mathcal{O}_2, \cdots, \mathcal{O}_n$ respectively.
At every time step $t$, an agent $i$ receives an observation $o_i^{(t)} \in \mathcal{O}_i$ and executes an action $a_i^{(t)} \in \mathcal{A}_i$ based on a stochastic policy $\pi^{(i)}: \mathcal{O}_i \times \mathcal{A}_i \rightarrow [0, 1]$.
Based on the executed action, the agent receives a reward $r_i^{(t)}: \mathcal{S} \times \mathcal{A}_i \rightarrow \mathbb{R}$ and the next observation $o_i^{(t+1)}$.
The state dynamics are determined by a transition function $\mathcal{T}: \mathcal{S} \times \mathcal{A}_1 \times \cdots \times \mathcal{A}_n \rightarrow \mathcal{S}$.
The agent policies are trained to maximize their own expected reward $\bar{r}_i = \sum_{t=1}^H r_i^{(t)}$ over a time horizon $H$.

\textbf{Extended Markov games.}
In this work, we are interested in interactions that involve not all but only a subset of agents.
For this purpose, we generalize Markov games as follows.
First, we augment the action space of each agent with a \texttt{NO-OP} (\textit{i.e.}, no action).
Then, we introduce a problem parameter, $2 \leq k \leq n$, with the following semantics.
During every rollout of the Markov game, all but $k$ agents deterministically execute the \texttt{NO-OP} operator while the $k$ agents execute actions as per the policies defined on the original observation and action spaces.
Accordingly, we assume that each agent receives rewards only in the interaction episode it participates in.
Informally, the extension allows for multiagent systems where all agents do not necessarily have to participate simultaneously in an interaction.
For instance, this allows to consider one-vs-one multiagent tournaments where only two players participate in any given match.

To further introduce the notation, consider a multiagent system as a generalized Markov game.
We denote the set of agent policies with $P = \{\pi^{(i)}\}_{i=1}^n$, interaction episodes with $E = \{E_{M_j}\}_{j=1}^m$ where $M_j \subseteq \{1, 2, \cdots, n\}, |M_j| = k$ is the set of $k$ agents participating in episode $E_{M_j}$.
To simplify presentation for the rest of the paper, we assume $k=2$ and, consequently, denote the set of interaction episodes between agents $i$ and $j$ as $E_{ij}$.
A single episode, $e_{ij}\in E_{ij}$, consists of a sequence of observations and actions for the specified time horizon, $H$.

\textbf{Imitation learning.}
Our approach to learning policy representations relies on behavioral cloning~\citep{pomerleau1991efficient}---a type of imitation learning where we train a mapping from observations to actions in a supervised manner.
Although there exist other imitation learning algorithms~\citep[\textit{e.g.}, inverse reinforcement learning,][]{abbeel2004apprenticeship}, our framework is largely agnostic to the choice of the algorithm, and we restrict our presentation to behavioral cloning, leaving other imitation learning paradigms to future work.

\section{Learning framework}

The dominant paradigm for unsupervised representation learning is to optimize the parameters of a representation function
that can best explain or generate the observed data.
For instance, the skip-gram objective used for language and graph data learns representations of words and nodes predictive of representations of surrounding context~\citep{mikolov2013distributed,grover2016node2vec}.
Similarly, autoencoding objectives, often used for image data, learn representations that can reconstruct the input~\citep{bengio2009learning}.

In this work, we wish to learn a representation function that maps episode(s) from an agent policy, $\pi^{(i)} \in \Pi$ to a real-valued vector embedding where $\Pi$ is a class of representable policies. That is, we optimize for the parameters $\theta$ for a function $f_\theta: \mathcal{E} \rightarrow \mathbb{R}^d$ where $\mathcal{E}$ denotes the space of episodes corresponding to a policy and $d$ is the dimension of the embedding.
Here, we have assumed the agent policies are black-boxes, \textit{i.e.}, we can only access them based on interaction episodes with other agents in a Markov game.
Hence, for every agent $i$, we wish to learn policies using $E_i = \cup_j E_{ij}^{(i)}$. Here, $E_{ij}^{(i)}$ refers the episode data for interactions between agent $i$ and $j$, but consisting of only the observation and action pairs of agent $i$.
For a multiagent system, we propose the following auxiliary tasks for learning a \textit{good representation} of an agent's policy:
\begin{enumerate}[topsep=0pt,itemsep=-2pt,partopsep=1ex,parsep=1ex]
\item \textit{Generative representations.} The representation should be useful for simulating the agent's policy.
\item \textit{Discriminative representations.} The representation should be able to distinguish the agent's policy with the policies of other agents.
\end{enumerate}
Accordingly, we now propose generative and discriminative objectives for representation learning in multiagent systems.

\subsection{Generative representations via imitation learning}

Imitation learning does not require direct access to the reward signal,
making it an attractive task for unsupervised representation learning.
Formally, we are interested in learning a policy $\pi^{(i)}_\phi: \mathcal{S} \times \mathcal{A} \rightarrow [0, 1]$ for an agent $i$ given access to observation and action pairs from interaction episode(s) involving the agent.
For behavioral cloning, we maximize the following (negative) cross-entropy objective:
\begin{align*}
\mathbb{E}_{e\sim E_i} \left[ \sum_{\langle o, a\rangle \sim e}\left[\log \pi_\phi^{(i)}(a \vert o)\right]\right]
\end{align*}
where the expectation is over interaction episodes of agent $i$ and the optimization is over the parameters $\phi$. 

Learning individual policies for every agent can be computationally and statistically prohibitive for large-scale multiagent systems, especially when the number of interaction episodes per agent is small. 
Moreover, it precludes generalization across the behaviors of such agents.
On the other hand, learning a single policy for all agents increases sample efficiency but comes at the cost of reduced modeling flexibility in simulating diverse agent behaviors. We offset this dichotomy by learning a single \textit{conditional} policy network.
To do so, we first specify a representation function, $f_\theta: \mathcal{E} \rightarrow \mathbb{R}^d$, with parameters $\theta$, where $\mathcal{E}$ represents the space of episodes. We use this embedding to condition the policy network.
Formally, the policy network is denoted by $\pi_{\phi, \theta}: \mathcal{S} \times \mathcal{A} \times \mathcal{E} \rightarrow [0, 1]$  and $\phi$ are parameters for the function mapping the agent observation and embedding to a distribution over the agent's actions.

The parameters $\theta$ and $\phi$ for the conditional policy network are learned jointly by maximizing the following objective:
\begin{align}\label{eq:conditional_imitation}
  \frac{1}{n} \sum_{i=1}^n \mathbb{E}_{\subalign{e_1 &\sim E_i, \\ e_2 &\sim E_i \backslash e_1}}\left[ \sum_{\langle o, a \rangle \sim e_1}\log \pi_{\phi, \theta}(a \vert o, e_2)\right]
\end{align}
For every agent, the objective function samples two distinct episodes $e_1$ and $e_2$. The observation and action pairs from $e_2$ are used to learn an embedding $f_\theta(e_2)$ that conditions the policy network trained on observation and action pairs from $e_1$.
The conditional policy network shares statistical strength through a common set of parameters for the policy network and the representation function across all agents. 

\begin{algorithm}[t]
  \caption{Learn Policy Embedding Function ($f_\theta$)}
  \label{alg:policyemb}
  \begin{algorithmic}[1]
  	\INPUT $\{E_i\}_{i=1}^n$ -- interaction episodes, $\lambda$ -- hyperparameter.
    \STATE Initialize $\theta$ and $\phi$
    \FOR{$i = 1, 2,\ldots, n$}
      \STATE Sample a positive episode $p_e \leftarrow e_+\sim E_i$
      \STATE Sample a reference episode $r_e \leftarrow e_*\sim E_i \backslash e_+$
      \STATE Compute $\mathrm{\texttt{Im\_loss}} \leftarrow - \sum\limits_{\langle o, a \rangle \sim e_+} \log \pi_{\phi, \theta}(a \vert o, e_*)$
      \FOR{$j = 1, 2,\ldots, n$}
        \IF {$j \neq i$}
          \STATE Sample a negative episode $n_e \leftarrow e_-\sim E_j$
          \STATE Compute $\mathrm{\texttt{Id\_loss}} \leftarrow d_\theta(e_+, e_-, e_*)$
          \STATE Set $\mathrm{\texttt{Loss}} \leftarrow \mathrm{\texttt{Im\_loss}} + \lambda \cdot \mathrm{\texttt{Id\_loss}}$
          \STATE Update $\theta$ and $\phi$ to minimize $\mathrm{\texttt{Loss}}$
        \ENDIF
      \ENDFOR
    \ENDFOR
    \OUTPUT $\theta$
  \end{algorithmic}
\end{algorithm}

\subsection{Discriminative representations via identification}

An intuitive requirement for any representation function learned for a multiagent system is that the embeddings should reflect characteristics of an agent's behavior that distinguish it from other agents.
To do so in an unsupervised manner, we propose an objective for \textit{agent identification}  based on the \textit{triplet loss} directly in the space of embeddings.

To learn a representation for agent $i$ based on interaction episodes, we use the representation function $f_\theta$ to compute three sets of embeddings: (i) a positive embedding for an episode $e_{+}\sim E_i$ involving agent $i$, (ii) a negative embedding for an episode $e_{-}\sim E_j$ involving a random agent $j\neq i$, and (iii) a reference embedding for an episode $e_{\ast}\sim E_i$ again involving agent $i$, but different from $e_{+}$.
Given these embeddings, we define the triplet loss:
\begin{equation}\label{eq:agent_id}
  \begin{split}
    \MoveEqLeft d_\theta(e_{+}, e_{-}, e_{\ast}) = \\
    & \left(1 + \exp\left\{\Vert r_e - n_e\Vert_2 - \Vert r_e - p_e \Vert_2\right\}\right)^{-2}
  \end{split}
\end{equation}

where $p_e = f_\theta(e_{+}), n_e = f_\theta(e_{-}), r_e = f_\theta(e_{\ast})$.
Intuitively, the loss encourages the positive embedding to be closer to the reference embedding than the negative embedding, which makes the embeddings of the same agent tend to cluster together and be further away from embeddings of other agents.
We note that various other notions of distance can also be used.
The one presented above corresponding to a squared softmax objective~\citep{hoffer2015deep}.

\subsection{Hybrid generative-discriminative representations}

Conditional imitation learning encourages $f_\theta$ to learn representations that can learn and simulate the entire policy of the agents and agent identification incentivizes representations that can distinguish between agent policies.
Both objectives are complementary, and we combine Eq.~\eqref{eq:conditional_imitation} and Eq.~\eqref{eq:agent_id} to get the final objective used for representation learning:
\begin{equation}\label{eq:hybrid_ob}
\frac{1}{n} \sum_{i=1}^n \mathbb{E}_{\subalign{e_+ &\sim E_i, \\ e_* &\sim E_i \backslash e_+}}\left[
	\begin{split}
    	&\underbrace{\sum_{\langle o, a \rangle \sim e_+}\log \pi_{\phi, \theta}(a \vert o, e_*)}_\text{imitation} - \\
        &\:\lambda \underbrace{\sum_{j \neq i}\mathbb{E}_{e_- \sim E_j}\left[d_\theta(e_{+}, e_{-}, e_{*}) \right]}_\text{agent identification}
    \end{split}
\right]
\end{equation}

where $\lambda>0$ is a tunable hyperparameter that controls the relative weights of the discriminative and generative terms.
The pseudocode for the proposed algorithm is given in Algorithm~\ref{alg:policyemb}. 
In experiments, we parameterize the conditional policy $\pi_{\theta,\phi}$ using neural networks and use stochastic gradient-based methods for optimization.

\begin{figure*}[t]
\centering
\begin{subfigure}[b]{0.33\textwidth}
\centering
 \begin{tikzpicture}
        \begin{scope}
            \Vertex[x=-2.25,y=1.25]{F}
            \Vertex[x=-1,y=2.25]{E}
            \Vertex[x=0,y=1.5]{A}
            \Vertex[x=-1.5,y=0]{G}
            \Vertex[x=0,y=0.25]{B}
            \Vertex[x=1.5,y=1.5]{C}
            \Vertex[x=1.5,y=0.25]{D}
        \end{scope}
        \draw[thick, black](A) -- (E);
        \draw[thick, black](G) -- (E);
        \draw[thick, black](F) -- (G);
        \draw[thick, black](G) -- (B);
        \draw[thick, black](F) -- (E);
       \draw[dashed, thick, red](B) -- (A);
       \draw[dotted, thick, green](C) -- (A);
       \draw[dotted, thick, green](D) -- (B);
       \draw[ dotted, thick, blue](D) -- (C);
    \end{tikzpicture}
\caption{Agent-Interaction Graph}
\label{fig:ai_graphs}
\end{subfigure}
\begin{subfigure}[b]{0.33\textwidth}
\centering
\includegraphics[width=0.85\columnwidth]{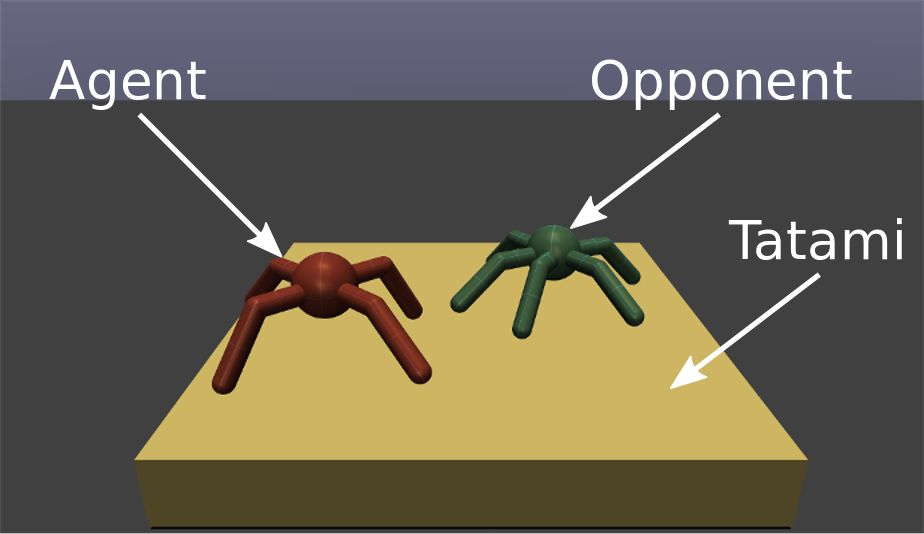}
\caption{The \texttt{RoboSumo} environment.}\label{fig:robosumo}
\end{subfigure}
\begin{subfigure}[b]{0.33\textwidth}
\centering
\includegraphics[width=0.95\columnwidth]{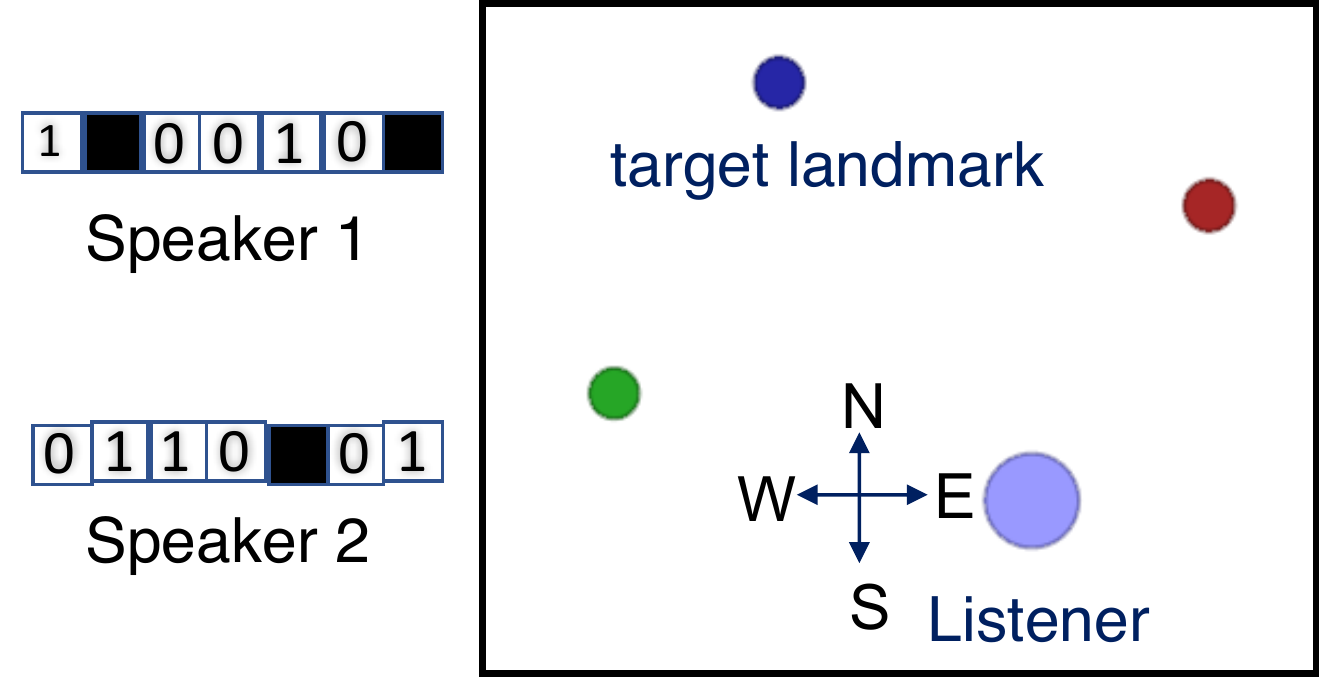}
\caption{The \texttt{ParticleWorld} environment.}\label{fig:particle}
\end{subfigure}
\caption{
An example of a graph used for evaluating generalization in a multiagent system (a).  Illustrations for the environments used in our experiments: competitive (b) and cooperative (c).
}
\end{figure*}
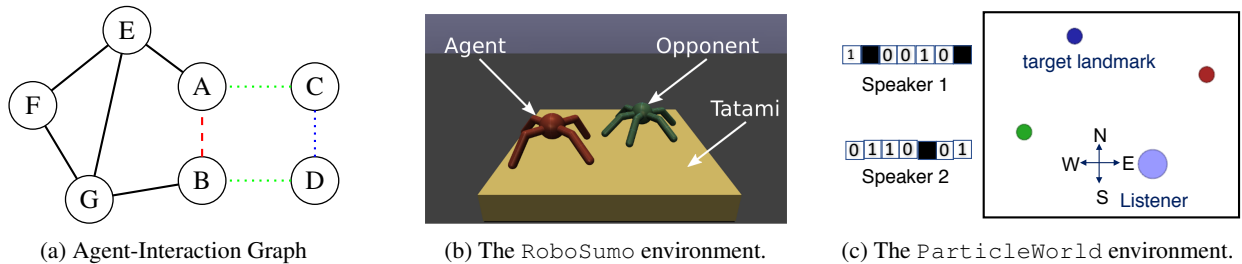

\section{Generalization in MAS}

Generalization is well-understood for supervised learning---models that shows similar train and test performance exhibit good generalization.
To measure the quality of the learned representations for a multiagent system (MAS), we introduce a graphical formalism for reasoning about agents and their interactions.

\subsection{Generalization across agents \& interactions}

In many scenarios, we are interested in generalization of the policy representation function $f_\theta$ across novel agents and interactions in a multiagent system.
For instance, we would like $f_\theta$ to output useful embeddings for a downstream task, even when evaluated with respect to unseen agents and interactions.
This notion of generalization is best understood using agent-interaction graphs~\citep{grover2018evaluating}.

The \textit{agent-interaction graph} describes interactions between a set of agent policies $P$ and a set of interaction episodes $I$ through a graph $G=(P, I)$.\footnote{If we have more than two participating agents per interaction episode, we could represent the interactions using a hypergraph.} An example graph is shown in Figure~\ref{fig:ai_graphs}. The graph represents a multiagent system consisting of interactions between pairs of agents, and we will especially focus on the interactions involving \textbf{A}lice, \textbf{B}ob, \textbf{C}harlie, and \textbf{D}avis. The interactions could be competitive (\textit{e.g.}, a match between two agents) or cooperative (\textit{e.g.}, two agents communicating for a navigation task).

We learn the representation function $f_\theta$ on a subset of the interactions, denoted by the solid black edges in Figure~\ref{fig:ai_graphs}. At test time, $f_\theta$ is evaluated on some downstream task of interest.
The agents and interactions observed at test time can be different from those used for training.
In particular, we consider the following cases:

\textbf{Weak generalization.}\footnote{Also referred to as \textit{intermediate generalization} by \citet{grover2018evaluating}.}
Here, we are interested in the generalization performance of the representation function on an unseen interaction between existing agents, all of which are observed during training. This corresponds to the red edge representing the interaction between \textbf{A}lice and \textbf{B}ob in Figure~\ref{fig:ai_graphs}. From the context of an agent-interaction graph, the test graph adds only edges to the train graph.

\textbf{Strong generalization.}
Generalization can also be evaluated with respect to unseen agents (and their interactions). This corresponds to the addition of agents \textbf{C}harlie and \textbf{D}avis in Figure~\ref{fig:ai_graphs}. Akin to a few shot learning setting, we observe a few of their interactions with existing agents \textbf{A}lice and \textbf{B}ob (green edges) and generalization is evaluated on unseen interactions involving \textbf{C}harlie and \textbf{D}avis (blue edges).  The test graph adds both nodes and edges to the train graph. 

For brevity, we skip discussion of weaker forms of generalization that involves evaluation of the test performance on unseen episodes of an existing training edge (black edge).

\subsection{Generalization across tasks}

Since the representation function is learned using an unsupervised auxiliary objective, we test its generalization performance by evaluating the usefulness of these embeddings for various kinds downstream tasks described below.

\textbf{Unsupervised.}
These embeddings can be used for clustering, visualization, and interpretability of agent policies in a low-dimensional space. Such semantic associations between the learned embeddings can be defined for a single agent wherein we expect representations for the same agent based on distinct episodes to be embedded close to each other, or across agents wherein agents with similar policies will have similar embeddings on average.

\textbf{Supervised.}
Deep neural network representations are especially effective for predictive modeling. In a multiagent setting, the embeddings serve as useful features for learning agent properties and interactions, including assignment of \textit{role} categories to agents with different skills in a collaborative setting, or prediction of win or loss outcomes of interaction matches between agents in a competitive setting.

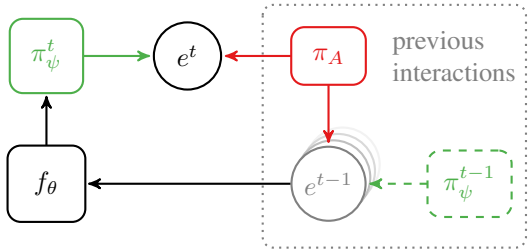
\begin{figure}[t]
\centering
\begin{tikzpicture}[->, >=stealth', shorten >=1pt, auto, node distance=1cm, thick, main/.style={draw, inner sep=8pt, outer sep=0pt, rounded corners=5pt}, x=1.25cm, y=1.7cm]
	\draw[dotted, draw=gray, rounded corners] (2.3, 0.4) rectangle (5.1, -1.5) {};
    \node[text width=1.8cm] (label) at (4.4, 0) {\color{gray} previous interactions};

	\node[main, color=nice-green] (pi_psi) at (0, 0) {$\pi^t_\psi$};
    \node[main, inner sep=10pt] (f_theta) at (0, -1) {$f_\theta$};
    \draw (f_theta) to (pi_psi);
    
    \node[circle, draw, inner sep=5pt] (e_curr) at (1.5, 0) {$e^t$};
    \node[main, color=nice-red] (pi_A) at (3.0, 0) {$\pi_A$};
    \draw[nice-green] (pi_psi) to (e_curr);
    \draw[nice-red] (pi_A) to (e_curr);
    
    \node[circle, fill=white, draw=gray!10, inner sep=3pt] (e_prev_1) at (3.15, -0.85) {\color{gray}$e^{t-1}$};
    \node[circle, fill=white, draw=gray!30, inner sep=3pt] (e_prev_2) at (3.10, -0.90) {\color{gray}$e^{t-1}$};
    \node[circle, fill=white, draw=gray!50, inner sep=3pt] (e_prev_3) at (3.05, -0.95) {\color{gray}$e^{t-1}$};
    \node[circle, fill=white, draw=gray, inner sep=3pt] (e_prev) at (3.00, -1.00) {\color{gray}$e^{t-1}$};
    \draw[nice-red] (pi_A) to (e_prev);
    \node[main, dashed, color=nice-green, inner sep=6pt] (pi_B) at (4.5, -1) {$\pi^{t-1}_\psi$};
    \draw[nice-green, dashed] (pi_B) to (e_prev);
    \draw (e_prev) to (f_theta);
\end{tikzpicture}
\caption{%
Illustration of the proposed model for optimizing a policy $\pi_\psi$ that conditions on an embedding of the opponent policy $\pi_A$.
At time $t$, the pre-trained representation function $f_\theta$ computes the opponent embedding based on a past interaction $e^{t-1}$.
We optimize $\pi_\psi$ to maximize the expected rewards in its current interactions $e^t$ with the opponent.
}
\label{fig:graphical_model_policy_opt}
\end{figure}

\textbf{Reinforcement.}
Finally, we can use the learned representation functions to improve generalization of the policies learned from a reinforcement signal in competitive and cooperative settings.
We design policy networks that, in addition to observations, take embedding vectors of the opposing agents as inputs.
The embeddings are computed from the past interactions of the opposing agent either with the agent being trained or with other agents using the representation function (Figure~\ref{fig:graphical_model_policy_opt}).
Such embeddings play the role of privileged information and allow us to train a policy network that uses this information to learn faster and generalize better to opponents or cooperators unseen at training time.

\begin{figure*}[t]
\centering
\begin{minipage}[b]{0.48\textwidth}
\begin{subfigure}[b]{0.48\columnwidth}
\centering
\includegraphics[width=\columnwidth]{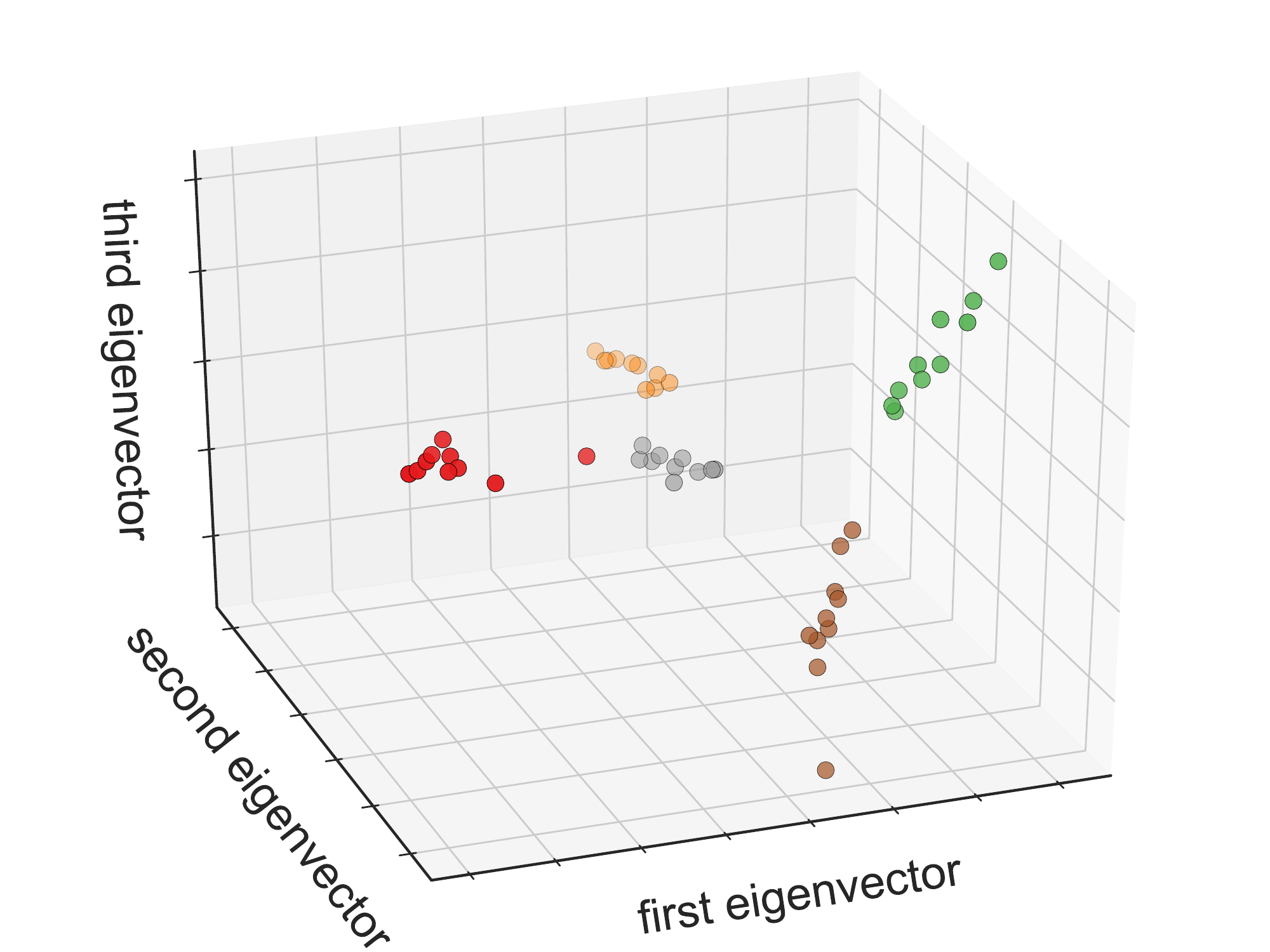}
\caption{\texttt{RoboSumo}: Weak}
\label{fig:comp_pca_weak}
\end{subfigure}
\begin{subfigure}[b]{0.48\columnwidth}
\centering
\includegraphics[width=\columnwidth]{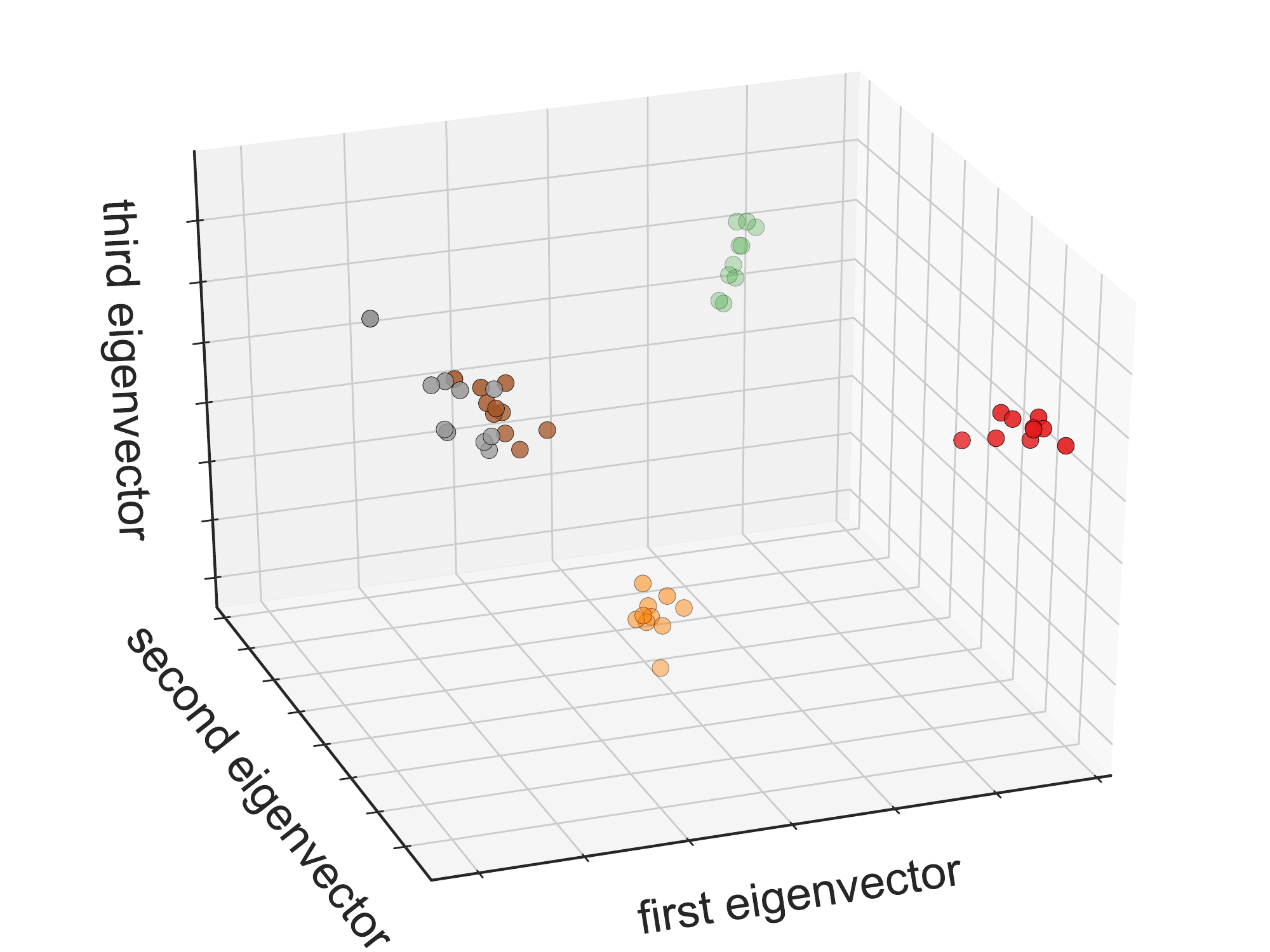}
\caption{\texttt{RoboSumo}: Strong}\label{fig:comp_pca_strong}
\end{subfigure}
\end{minipage}
\begin{minipage}[b]{0.48\textwidth}
\centering
\begin{subfigure}[b]{0.48\columnwidth}
\centering
\includegraphics[width=\columnwidth]{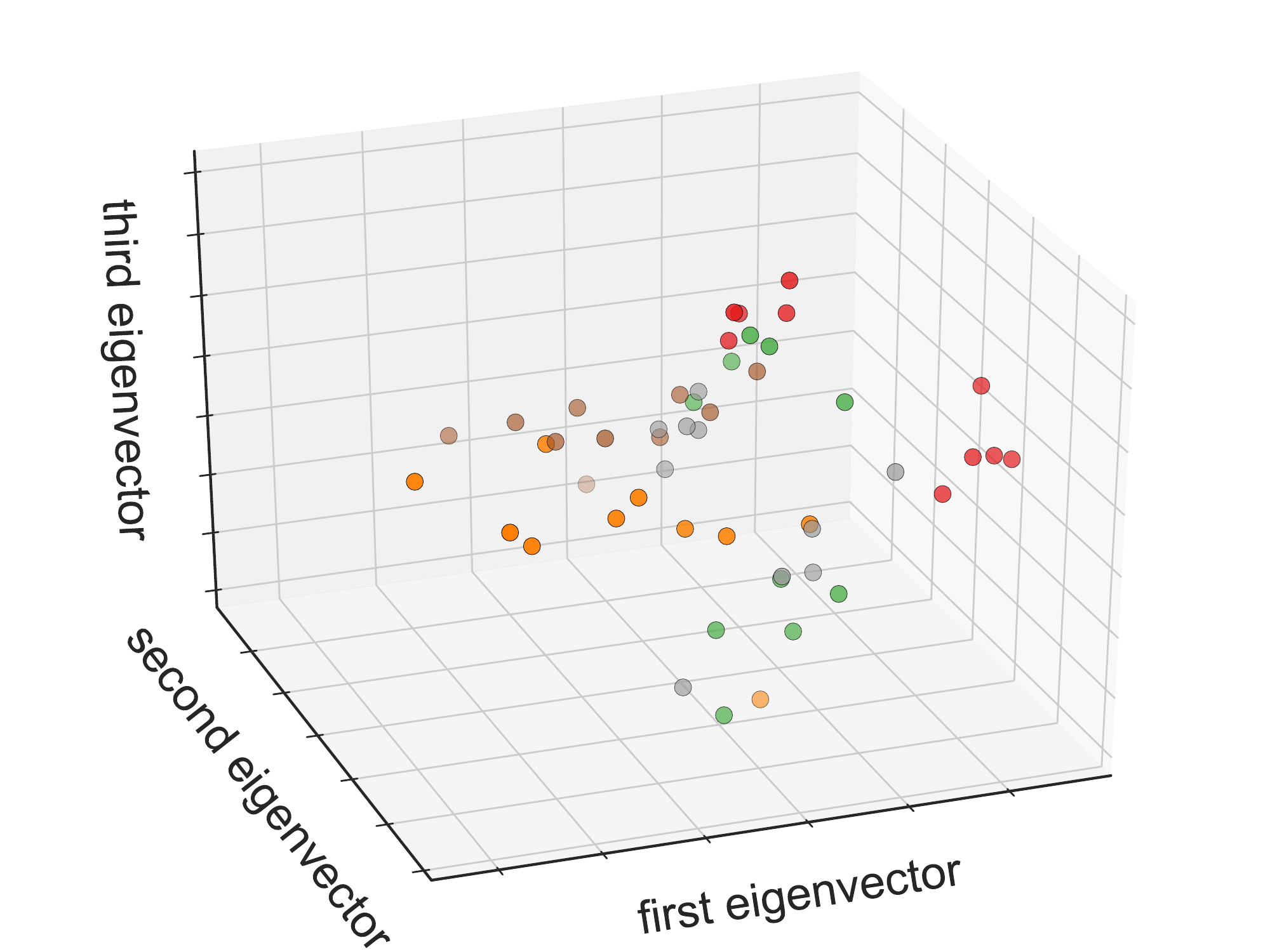}
\caption{\texttt{ParticleWorld}: Weak}
\label{fig:coop_pca_weak}
\end{subfigure}
\begin{subfigure}[b]{0.48\columnwidth}
\centering
\includegraphics[width=\columnwidth]{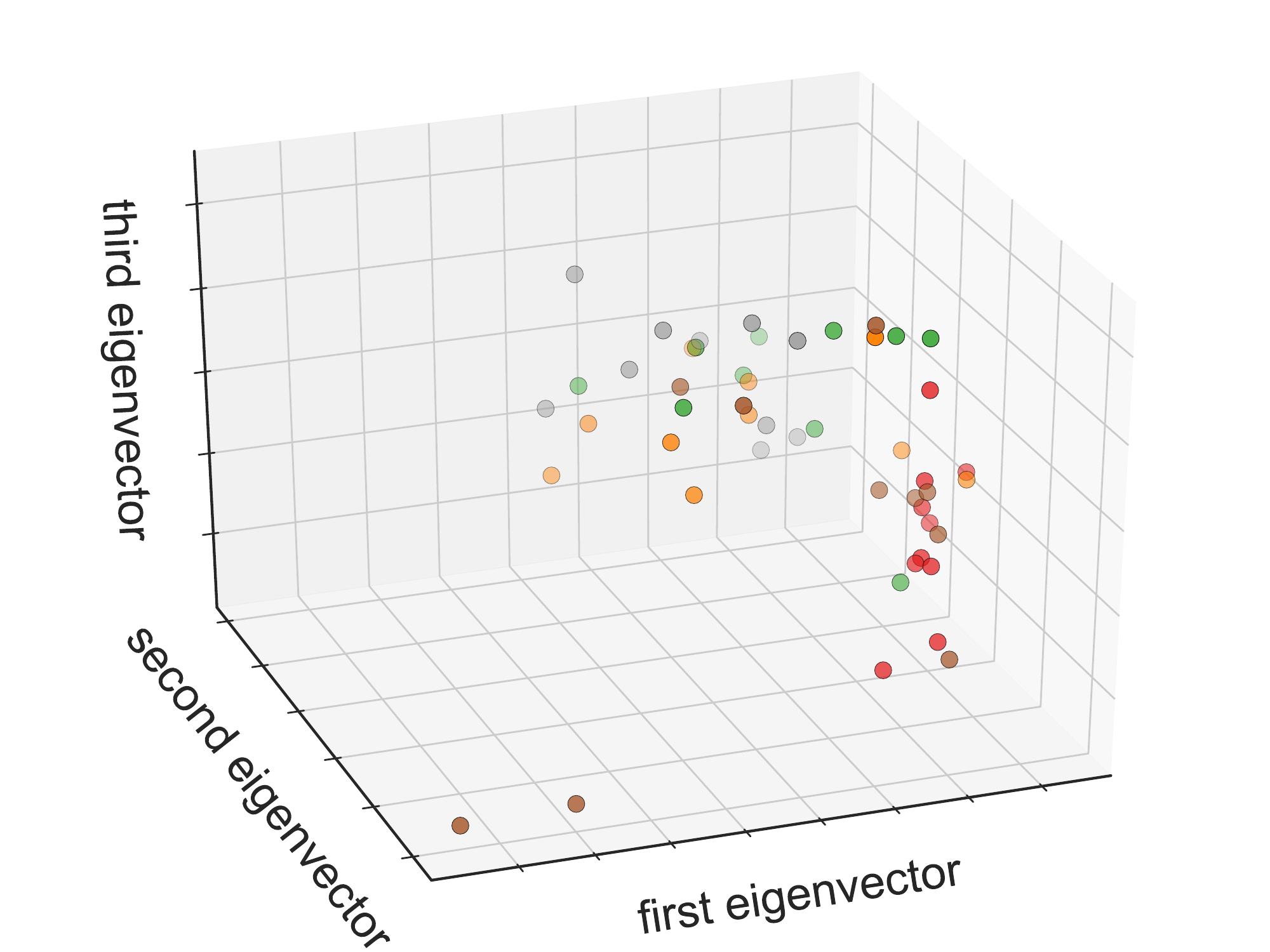}
\caption{\texttt{ParticleWorld}: Strong}\label{fig:coop_pca_strong}
\end{subfigure}
\end{minipage}
\caption{Embeddings learned using \texttt{Emb-Hyb} for $10$ test interaction episodes of $5$ agents projected on the first three principal components for \texttt{RoboSumo}  and \texttt{ParticleWorld}. Color denotes agent policy.}\label{fig:comp_pca}
\end{figure*}

\section{Evaluation methodology \& results}

We evaluate the proposed framework for both competitive and collaborative environments on various downstream machine learning tasks.
In particular, we use the \texttt{RoboSumo} and \texttt{ParticleWorld} environments for the competitive and collaborative scenarios, respectively. We consider the embedding objectives in Eq.~\eqref{eq:conditional_imitation}, Eq.~\eqref{eq:agent_id}, and Eq.~\eqref{eq:hybrid_ob} independently and refer to them as \texttt{Emb-Im}, \texttt{Emb-Id}, and \texttt{Emb-Hyb} respectively. The hyperparameter $\lambda$ for \texttt{Emb-Hyb} is chosen by grid search over $\lambda \in \{0.01, 0.05, 0.1, 0.5\}$ on a held-out set of interactions. 

In all our experiments, the representation function $f_\theta$ is specified through a multi-layer perceptron (MLP) that takes as input an episode and outputs an embedding of that episode. In particular, the MLP takes as input a single (observation, action) pair to output an intermediate embedding. We average the intermediate embeddings for all (observation, action) pairs in an episode to output an episode embedding. To condition a policy network on the embedding, we simply concatenate the observation fed as input to the network with the embedding. 
Experimental setup and other details beyond what we state below are deferred to the Appendix.

\subsection{The \texttt{RoboSumo} environment}

For the competitive environment, we use \texttt{RoboSumo}~\citep{al2017continuous}---a 3D environment with simulated physics (based on MuJoCo~\citep{todorov2012mujoco}) that allows agents to control multi-legged 3D robots and compete against each other in continuous-time wrestling games (Figure~\ref{fig:robosumo}). For our analysis, we train a diverse collection of $25$ agents, some of which are trained via self-play and others are trained in pairs concurrently using Proximal Policy Optimization (PPO) algorithm~\citep{schulman2017ppo}.

We start with a fully connected agent-interaction graph (clique) of $25$ agents. Every edge in this graph corresponds to $10$ rollout episodes involving the corresponding agents. The maximum length (or horizon) of any episode is $500$ time steps, after which the episode is declared a draw.
To evaluate weak generalization, we sample a connected subgraph for training with approximately $60\%$ of the edges preserved for training, and remaining split equally for validation and testing. 
For strong generalization, we preserve $15$ agents and their interactions with each other for training, and similarly, $5$ agents and their within-group interactions each for validation and testing.
\begin{table}[t]
\centering
\caption{Intra-inter clustering ratios (IICR) and accuracies for outcome prediction (Acc) for weak (W) and strong (S) generalization on \texttt{RoboSumo}. }\label{tab:comp_unsup_sup}
\begin{tabular}{l|cc|cc}                                                                & IICR (W)          & IICR (S)& Acc (W)& Acc(S)\\\hline
\texttt{Emb-Im}                                                       & $0.24$          & $0.23$ & $0.71$          & $\mathbf{0.60 }$   \\
\texttt{Emb-Id}                                                       & $0.25$          & $0.27$ & $0.67$          & $0.56$   \\
\texttt{Emb-Hyb} & $\mathbf{0.22}$ & $\mathbf{0.21}$  & $\mathbf{0.73}$ & $0.56$ 
\end{tabular}
\vspace{-4ex}
\end{table}

\begin{figure*}[t]
\centering
\includegraphics[width=0.5\textwidth]{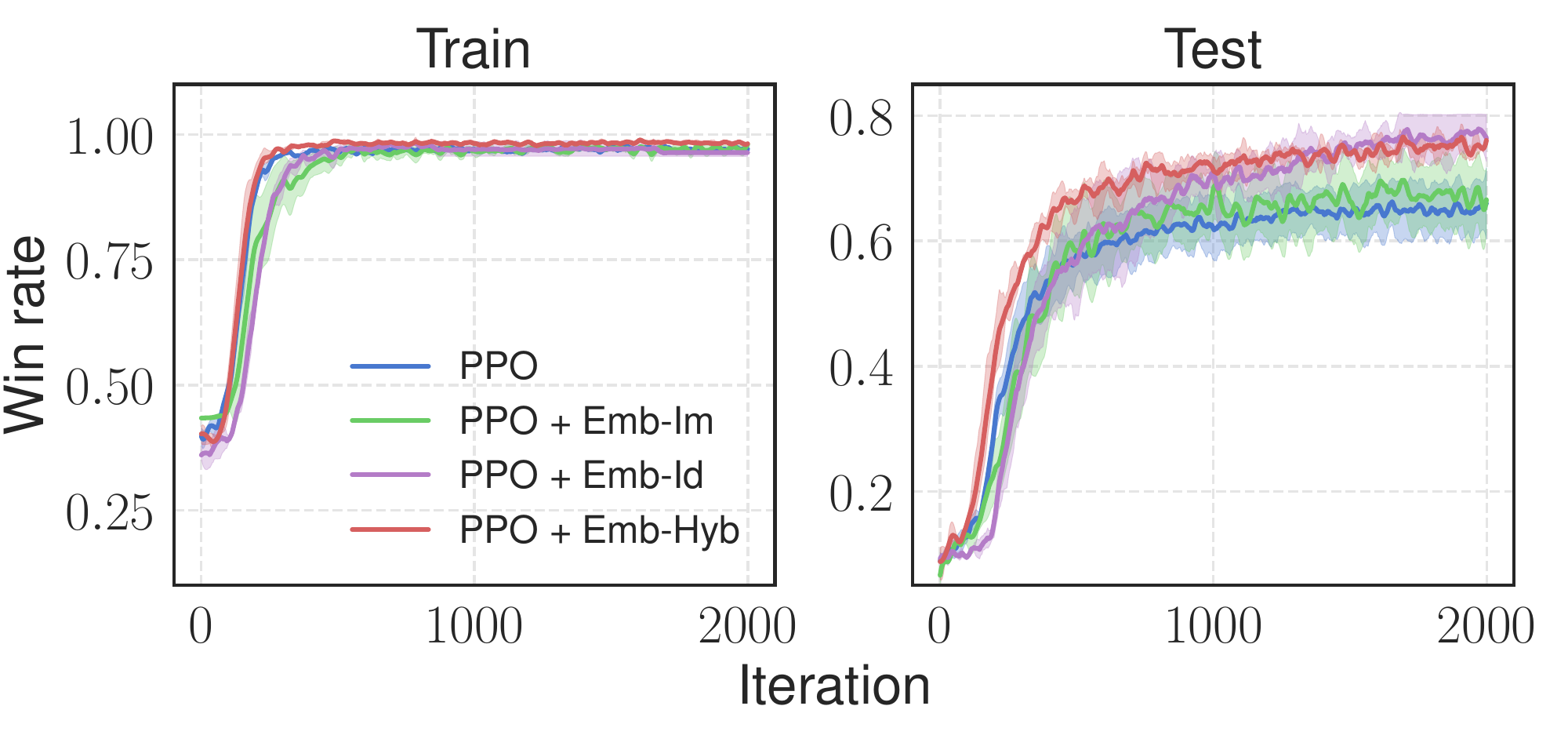}%
\includegraphics[width=0.5\textwidth]{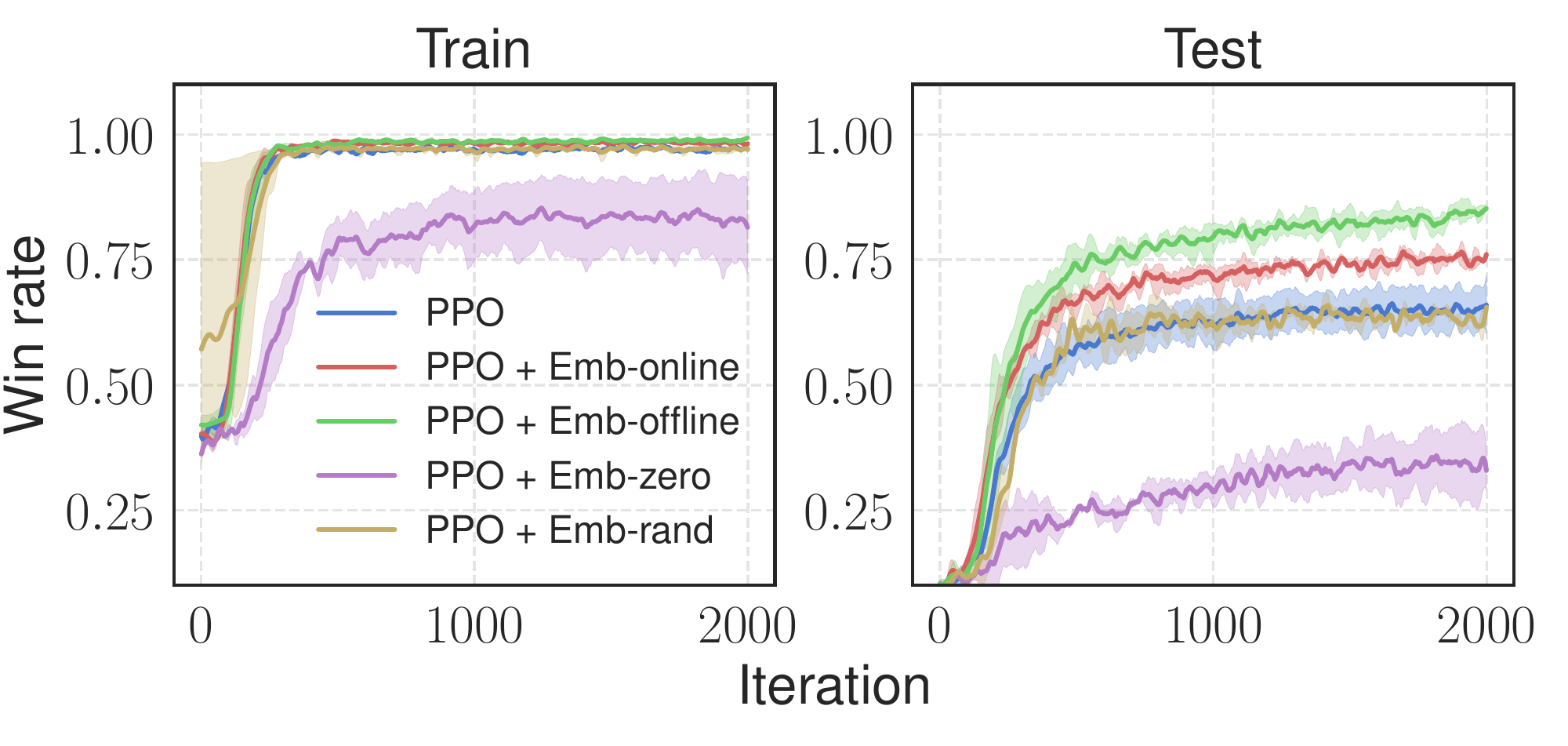}
\vspace{-2em}
\caption{%
Average win rates of the newly trained agents against $5$ training agent and $5$ testing agents.
The left two charts compare baseline with policies that make use of \texttt{Emb-Im}, \texttt{Emb-Id}, and \texttt{Emb-Hyb} (all computed online).
The right two charts compare different embeddings used at evaluation time (all embedding-conditioned policies use \texttt{Emb-Hyb}).
At each iteration, win rates were computed based on 50 1-on-1 games.
Each agent was trained 3 times, each time from a different random initialization.
Shaded regions correspond to 95\% CI.
}\label{fig:competitive-lvl2-plots}
\end{figure*}

\begin{figure*}[t]
\centering
\includegraphics[width=\textwidth]{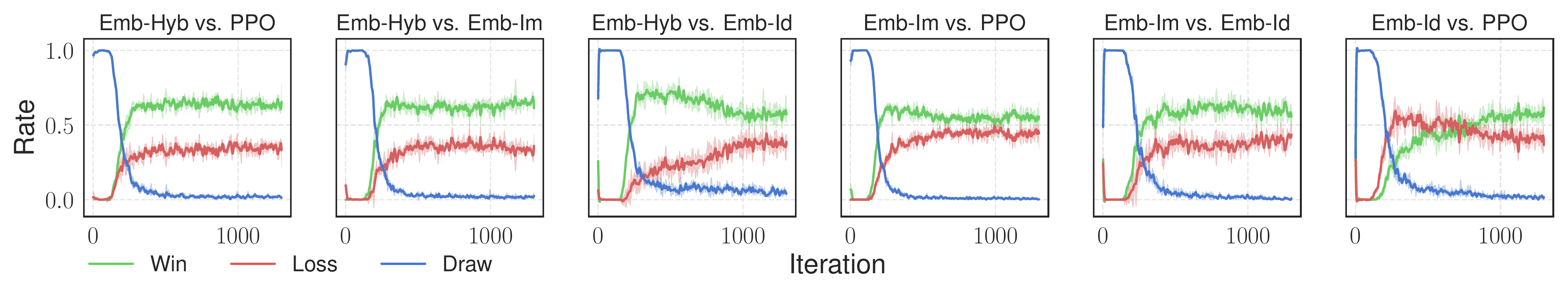}
\caption{%
Win, loss, and draw rates plotted for the first agent in each pair.
Each pair of agents was evaluated after each training iteration on 50 1-on-1 games; curves are based on 5 evaluation runs.
Shaded regions correspond to 95\% CI.
}\label{fig:competitive-lvl3-plots}
\end{figure*}

\begin{figure}[t]
\centering
\vspace{-2ex}
\includegraphics[width=0.75\columnwidth]{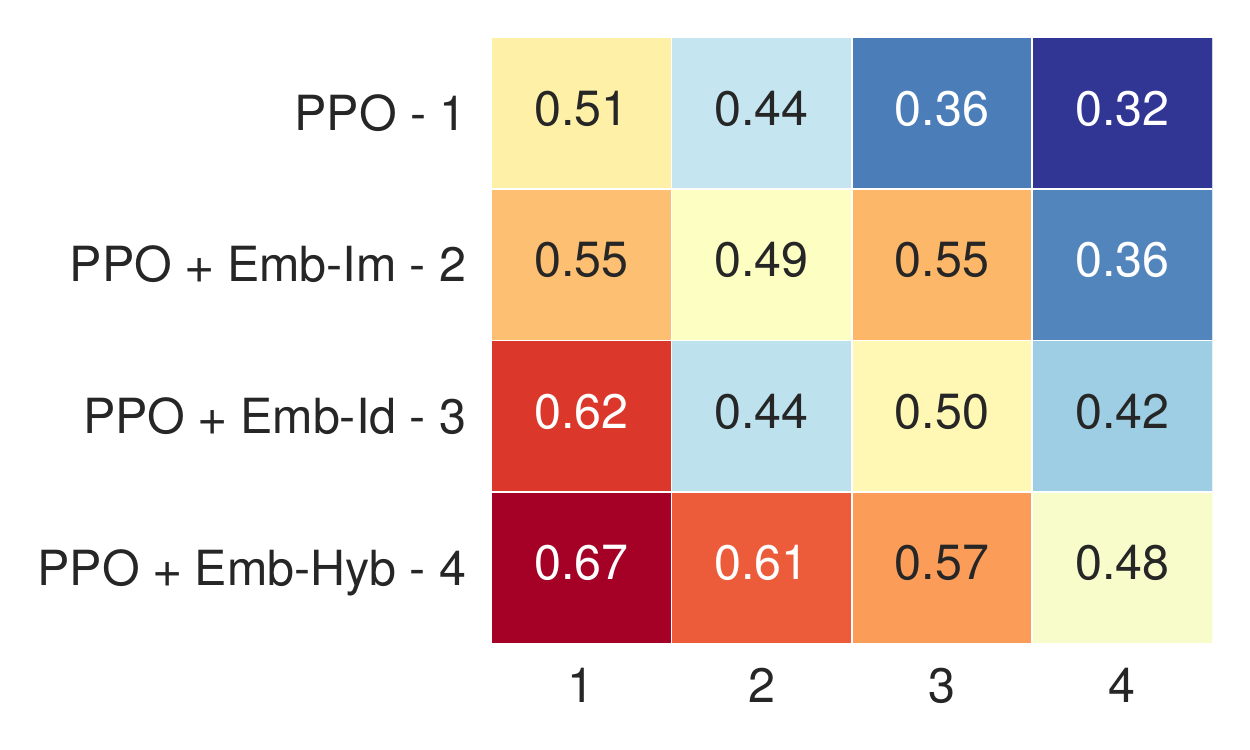}
\vspace{-1ex}
\caption{%
Win rates for agents specified in each row at computed at iteration 1000.}\label{fig:competitive-lvl3-heatmap}
\vspace{-2ex}
\end{figure}

\subsubsection{Embedding analysis}
To evaluate the robustness of the embeddings, we compute multiple embeddings for each policy based on different episodes of interaction at test time.
Our evaluation metric is based on the intra- and inter-cluster Euclidean distances between embeddings. 
The intra-cluster distance for an agent is the average pairwise distance between its embeddings computed on the set of test interaction episodes involving the agent.
Similarly, the inter-cluster distance is the average pairwise distance between the embeddings of an agent with those of other agents. 
Let $T_i = \{t^{(i)}_c\}_{c=1}^{n_i}$ denote the set of test interactions involving agent $i$. 
We define the intra-inter cluster ratio (IICR) as:
\begin{align*}
\mathrm{IICR} &= \frac{\frac{1}{n}\sum_{i=1}^n \frac{1}{n_i^2}\sum_{a=1}^{n_i}\sum_{b=1}^{n_i}\Vert t^{(i)}_a - t^{(i)}_b \Vert_2}{\frac{1}{n(n-1)}\sum\limits_{i=1}^n \sum\limits_{j \neq i}^n \frac{1}{n_i n_j}\sum\limits_{a=1}^{n_i}\sum\limits_{b=1}^{n_j}\Vert t^{(i)}_a - t^{(j)}_b \Vert_2}.
\end{align*}

The intra-inter clustering ratios are reported in Table~\ref{tab:comp_unsup_sup}.
A ratio less than $1$ suggests that there is signal that identifies the agent, and the signal is stronger for lower ratios.
Even though this task might seem especially suited for the agent identification objective, we interestingly find that the \texttt{Emb-Im} attains lower clustering ratios than \texttt{Emb-Id} for both weak and strong generalization.
\texttt{Emb-Hyb} outperforms both these methods.
We qualitatively visualize the embeddings learned using \texttt{Emb-Hyb} by projecting them on the leading principal components, as shown in Figures~\ref{fig:comp_pca_weak} and~\ref{fig:comp_pca_strong} for $10$ test interaction episodes of $5$ randomly selected agents in the weak and strong generalization settings respectively.

\subsubsection{Outcome prediction}
We can use these embeddings directly for training a classifier to predict the outcome of an episode (win/loss/draw).
For classification, we use an MLP with $3$ hidden layers of $100$ units each and the learning objective minimizes the cross entropy error. The input to the classifier are the embeddings of the two agents involved in the episode.
The results are reported in Table~\ref{tab:comp_unsup_sup}.
Again, imitation based methods seem more suited for this task with \texttt{Emb-Hyb} and \texttt{Emb-Im} outperforming other methods for weak and strong generalization respectively.

\subsubsection{Policy optimization}
Here we ask whether embeddings can be used to improve learned policies in a reinforcement learning setting both in terms of end performance and generalization.
To this end, we select 5 training, 5 validation, and 5 testing opponents from the pool of 25 pre-trained agents.
Next, we train a new agent with reinforcement learning to compete against the selected 5 training opponents; the agent is trained concurrently against all 5 opponents using a distributed version of PPO algorithm, as described in~\citet{al2017continuous}.
Throughout training, we evaluate new agents on the 5 testing opponents and record the average 
win and draw rates.

Using this setup, we compare a baseline agent with MLP-based policy
with an agent whose policy takes 100-dimensional embeddings of the opponents as additional inputs at each time step
and uses that information to condition its behavior on the opponent's representation.
The embeddings for each opponent are either computed \textit{online}, \textit{i.e.}, based on an interaction episode rolled out during training at a previous time step (Figure~\ref{fig:graphical_model_policy_opt}), or \textit{offline}, \textit{i.e.}, pre-computed before training the new agent using only interactions between the pre-trained opponents.

Figure~\ref{fig:competitive-lvl2-plots} shows the average win rates against the set of training and testing opponents for the baseline and our agents that use different types of embeddings.
While every new agent is able to achieve almost 100\% win rate against the training opponents, we see that the agents that condition their policies on the opponent's embeddings perform better on the held-out set of opponents, \textit{i.e.}, generalize better, with the best performance achieved with \texttt{Emb-Hyb}.
We also note that embeddings computed offline turn out to lead to better performance than if computed online\footnote{Perhaps, this is due to differences in the interactions of the opponents between themselves and with the new agent that the embedding network was not able to capture entirely.}.
As an ablation test, we also evaluate our agents when they are provided an incorrect embedding (either all zeros, \texttt{Emb-zero}, or an embedding selected for a different random opponent, \texttt{Emb-rand}) and observe that such embeddings lead to a degradation in performance\footnote{Performance decrease is most significant for \texttt{Emb-zero}, which is an out-of-distribution all-zeros vector.}.

Finally, to evaluate strong generalization in the RL setting, we pit the newly trained baseline and agents with embedding-conditional policies against each other.
Since the embedding network has never seen the new agents, it must exhibit strong generalization to be useful in such setting.
The results are give in Figures~\ref{fig:competitive-lvl3-plots} and~\ref{fig:competitive-lvl3-heatmap}.
Even though the margin is not very large, the agents that use \texttt{Emb-Hyb} perform the best on average.

\subsection{The \texttt{ParticleWorld} environment}

For the collaborative setting, we evaluate the framework on the \texttt{ParticleWorld} environment for cooperative communication~\citep{mordatch2017emergence,lowe2017multi}. The environment consists of a continuous $2$D grid with $3$ landmarks and two kinds of agents collaborating to navigate to a common landmark goal (Figure~\ref{fig:particle}). At the beginning of every episode, the \textit{speaker} agent is shown the RGB color of a single target landmark on the grid. The speaker then communicates a fixed length binary message to the \textit{listener} agent. Based on the received messages, the listener agent the moves in a particular direction. The final reward, shared across the speaker and listener agents, is the distance of the listener to the target landmark after a fixed time horizon. 

The agent-interaction graph for this environment is bipartite with only cross edges between speaker and listener agents. Every interaction edge in this graph corresponds to $1000$ rollout episodes where the maximum length of any episode is $25$ steps. 
We pretrain $28$ MLP parameterized speaker and listener agent policies.
Every speaker learns through communication with only two different listeners and vice-versa, giving an extremely sparse agent-interaction graph. We explicitly encoded diversity in these speakers and listener agents by masking bits in the communication channel. 
In particular, we masked $1$ or $2$ randomly selected bits for every speaker agent in the graph to give a total of $\binom{7}{1} + \binom{7}{2}=28$ distinct speaker agents. 
Depending on the neighboring speaker agents in the agent-interaction graph, the listener agents also show diversity in the learned policies. The policies are learned using multiagent deep deterministic policy gradients~\citep[MADDPG,][]{lowe2017multi}.

\begin{table}[t]
\centering
\caption{Intra-inter clustering ratios (IICR) for weak (W) and strong (S) generalization on \texttt{ParticleWorld}. Lower is better.}\label{tab:coop_clustering_ratio}
\vspace{-1em}
\begin{tabular}{l|cc}                                                                & IICR (W)          & IICR (S) \\\hline
\texttt{Emb-Im}   & $0.58$          & $0.86$   \\
\texttt{Emb-Id}    & $\mathbf{0.50 }$         & $\mathbf{0.82}$   \\
\texttt{Emb-Hyb} & $0.54$ & $0.85$  
\end{tabular}
\end{table}

\begin{table}[t]
  \centering
  \caption{Average train and test rewards for speaker policies on \texttt{ParticleWorld}. }\label{tab:coop_weak_gen}
  \vspace{-1em}
  \begin{tabular}{l|cc}                                                                & Train    &  Test \\\hline
    MADDPG & $\mathbf{-11.66}$ & $-18.99$  \\
    MADDPG + \texttt{Emb-Im} & $-11.68$ & $-17.75$  \\
    MADDPG + \texttt{Emb-Id} & $-11.68$ & $-17.68$  \\
    MADDPG + \texttt{Emb-Hyb} & $-11.77$ & $\mathbf{-17.20}$  \\   
  \end{tabular}
  \vspace{-1em}
\end{table}

In this environment, the speakers and listeners are tightly coupled. Hence we vary the setup used previously in the competitive scenario. We wish to learn embeddings of listeners based on their interactions with speakers. Since the agent-interaction graph is bipartite, we use the embeddings of listener agents to condition a shared policy network for the respective speaker agents.

\subsubsection{Embedding analysis}

For the weak generalization setting, we remove an outgoing edge from every listener agent in the original graph to obtain the training graph. In the case of strong generalization, we set aside $7$ listener agents (and their outgoing edges) each for validation and testing while the representation function is learned on the remaining 14  listener agents and their interactions. The intra-inter clustering ratios are shown in Table~\ref{tab:coop_clustering_ratio},
and the projections of the embeddings learned using \texttt{Emb-Hyb} are visualized in Figure~\ref{fig:coop_pca_weak} and Figure~\ref{fig:coop_pca_strong} for weak and strong generalization respectively.
In spite of the high degree of sparsity in the training graph, the intra-inter clustering ratio for the test interaction embeddings is less than unity suggesting an agent-specific signal.
\texttt{Emb-id} works particularly well in this environment, achieving best results for both weak and strong generalization.

\subsubsection{Policy optimization}
Here, we are interested in learning speaker agents that can communicate more effectively with a much wider range of listeners given knowledge of their embeddings. Referring back to Figure~\ref{fig:graphical_model_policy_opt}, we learn a policy $\pi_\psi$ for a speaker agent that conditions on the representation function $f_\theta$ for the listener agents. For cooperative communication, we consider interactions with 14 pre-trained listener agents split as $6$ training, $4$ validation, and $4$ test agents.\footnote{None of the methods considered were able to learn a non-trivial speaker agent when trained simultaneously with all $28$ listener agents. Hence, we simplified the problem by considering the $14$ listener agents that attained the best rewards during pretraining.} Similar to the competitive setting, we compare performance against a baseline speaker agent that does not have access to any privilege information about the listeners. 
We summarize the results for the best validated models during training and $100$ interaction episodes per test listener agent across $5$ initializations in Table~\ref{tab:coop_weak_gen}. From the results, we observe that online embedding based methods can generalize better than the baseline methods. The baseline MADDPG achieves the lowest training error, but fails to generalize well enough and incurs a low average reward for the test listener agents.

\section{Discussion \& Related Work}\label{sec:related}

Agent modeling is a well-studied topic within multiagent systems. See \citet{albrecht2017autonomous} for an excellent recent survey on this subject. The vast majority of literature concerns with learning models for a specific predictive task. Predictive tasks are typically defined over actions, goals, and beliefs of other agents~\citep{stone2000multiagent}. In competitive domains such as Poker and Go, such tasks are often integrated with domain-specific heuristics to model opponents and learn superior policies~\citep{rubin2011computer,mnih2015human}. Similarly, intelligent tutoring systems take into account pedagogical features of students and teachers to accelerate learning of desired behaviors in a collaborative environment~\citep{mccalla2000active}. 

In this work, we proposed an approach for modeling agent behavior in multiagent systems through unsupervised representational learning of agent policies.
Since we sidestep any domain specific assumptions and learn in an unsupervised manner, our framework learns representations that are useful for several downstream tasks.
This extends the use of deep neural networks in multiagent systems to applications beyond traditional reinforcement learning and predictive modeling~\citep{mnih2015human,hoshen2017vain}.

Both the generative and discriminative components of our framework have been explored independently in prior work.
Imitation learning has been extensively studied in the single-agent setting and recent work by \citet{le2017coordinated} proposes an algorithm for imitation in a coordinated multiagent system.
\citet{wang2017robust} proposed an imitation learning algorithm for learning robust controllers with few expert demonstrations in a single-agent setting that conditions the policy network on an inference network, similar to the encoder in our framework.
In another recent work, \citet{li2017inferring} propose an algorithm for learning interpretable representations using generative adversarial imitation learning. 
Agent identification which represents the discriminative term in the learning objective is inspired from triplet losses and Siamese networks that are used for learning representations of data using distance comparisons~\citep{hoffer2015deep}.

A key contribution of this work is a principled methodology for evaluating generalization of representations in multiagent systems based on the graphs of the agent interactions.
Graphs are a fundamental abstraction for modeling relational data, such as the interactions arising in multiagent systems~\citep{zhou2016dynamics,zhou2016game,chen2017multiplayer,battaglia2016interaction,hoshen2017vain} and concurrent work proposes to learn such graphs directly from data~\citep{kipf2018neural}. 
\section{Conclusion \& Future Work}

In this work, we presented a framework for learning representations of agent policies in multiagent systems. The agent policies are accessed using a few interaction episodes with other agents. Our learning objective is based on a novel combination of a generative component based on imitation learning and a discriminative component for distinguishing the embeddings of different agent policies. Our overall framework is unsupervised, sample-efficient, and domain-agnostic, and hence can be readily extended to many environments and downstream tasks. Most importantly, we showed the role of these embeddings as privileged information for learning more adaptive agent policies in both collaborative and competitive settings.

In the future, we would like to explore multiagent systems with more than two agents participating in the interactions.
Semantic interpolation of policies directly in the embedded space in order to obtain a policy with desired behaviors quickly is another promising direction.
Finally, it would be interesting to extend and evaluate the proposed framework to learn representations for history dependent policies such as those parameterized by long short-term memory networks. 
\section*{Acknowledgements}
We are thankful to Lisa Lee, Daniel Levy, Jiaming Song, and everyone at OpenAI for helpful comments and discussion.
AG is supported by a Microsoft Research PhD Fellowship.
MA is partially supported by NIH R01GM114311.
JKG is partially supported by the Army Research Laboratory through the Army High Performance Computing Research Center under Cooperative Agreement W911NF-07-2-0027.

\bibliography{references}
\balance
\bibliographystyle{icml2018} 
\pagebreak

\appendix

\section{Experimental Setup}
\subsection*{\texttt{RoboSumo} Environment}
To limit the scope of our study, we restrict agent morphologies to only $4$-leg robots.
During the game, observations of each agent were represented by a 120-dimensional vector comprised of positions and velocities of its own body and positions of the opponent's body; agent's actions were 8-dimensional vectors that represented torques applied to the corresponding joints.

\subsubsection*{Network Architecture}
Agent policies are parameterized as multi-layer perceptrons (MLPs) with 2 hidden layers of 90 units each.
For the embedding network, we used another MLP network with 2 hidden layers of 100 units each to give an embedding of size $100$. For the conditioned policy network we also reduce the hidden layer size to 64 units each.

\subsubsection*{Policy Optimization}
For learning the population of agents, we use the distributed version of PPO algorithm as described in~\citep{al2017continuous} with $2\times10^{-3}$ learning rate, $\epsilon = 0.2$, 16,000 time steps per update with 6 epochs 4,000 time steps per batch.

\subsubsection*{Training}
For our analysis, we train a diverse collection of 25 agents, some of which are trained via self-play and others are trained in pairs concurrently, forming a clique agent-interaction graph.

\begin{figure}[h]
\centering
\begin{tikzpicture}
  \graph[nodes={draw, circle, thick, fill=nice-green}, clique, n=10, clockwise, radius=3cm]
  {
    1/"$1$", 2/"$2$", 3/"$3$", 4/"$4$", 5/"$5$", 6/"$6$", 7/"$7$", 8/"$8$", 9/"$9$", 10/"$10$"
  };
\end{tikzpicture}
\caption{An example clique agent interaction graph with $10$ agents.}
\end{figure}
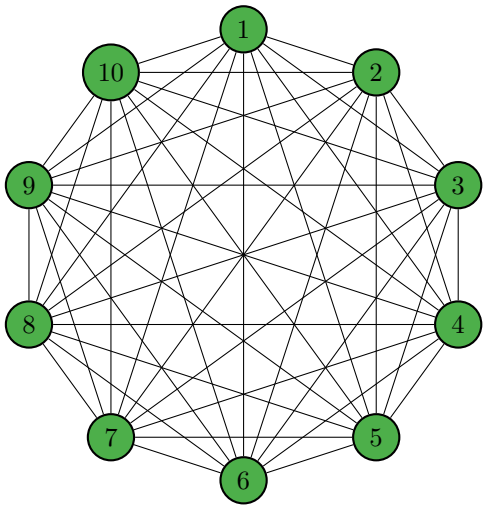

\subsection*{\texttt{ParticleWorld} Environment}
The overall continuous observation and discrete action space for the speaker agents are $3$ and $7$ dimensions respectively. For the listener agents, the observation and action spaces are $15$ and $5$ dimensions respectively.

\subsubsection*{Network Architecture}
Agent policies and shared critic (\textit{i.e.}, a value function) are parameterized as multi-layer perceptrons (MLPs) with 2 hidden layers of 64 units each.
The observation space for the speaker is small ($3$ dimensions), and a small embedding of size $5$ for the listener policy gives good performance. For the embedding network, we again used an MLP with $2$ hidden layers of $100$ units each.

\subsubsection*{Policy Optimization}
For learning the initial population of listener and agent policies, we use multiagent deep deterministic policy gradients (MADDPG) as the base algorithm~\citep{lowe2017multi}. Adam optimizer~\citep{kingma2014adam} with a learning rate of $4\times10^{-3}$ was used for optimization. Replay buffer size was set to $10^6$ timesteps.

\subsubsection*{Training}
We first train $28$ speaker-listener pairs using the MADDPG algorithm. From this collection of $28$ speakers, we train another set of $28$ listeners, each trained to work with a speaker pair, forming a bipartite agent-interaction graph. We choose the best $14$ listeners for later experiments.

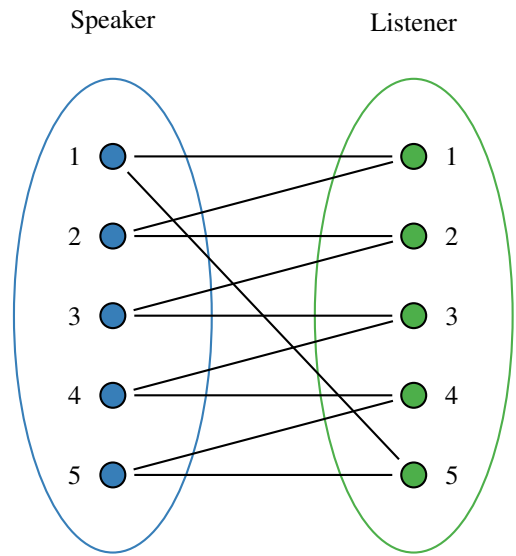
\begin{figure}[h]
\centering
\begin{tikzpicture}[thick,
  every node/.style={draw,circle},
  fsnode/.style={fill=nice-blue},
  ssnode/.style={fill=nice-green},
  every fit/.style={ellipse,draw,inner sep=-2pt,text width=2cm},
  shorten >= 3pt,shorten <= 3pt
]

\begin{scope}[start chain=going below,node distance=7mm]
\foreach \i in {1,2,...,5}
  \node[fsnode,on chain] (f\i) [label=left: \i] {};
\end{scope}

\begin{scope}[xshift=4cm,start chain=going below,node distance=7mm]
\foreach \i in {1,2,...,5}
  \node[ssnode,on chain] (s\i) [label=right: \i] {};
\end{scope}

\node [nice-blue,fit=(f1) (f5),label=above:Speaker] {};
\node [nice-green,fit=(s1) (s5),label=above:Listener] {};

\draw (f1) -- (s1);
\draw (f2) -- (s1);
\draw (f2) -- (s2);
\draw (f3) -- (s2);
\draw (f3) -- (s3);
\draw (f4) -- (s3);
\draw (f4) -- (s4);
\draw (f5) -- (s4);
\draw (f5) -- (s5);
\draw (f1) -- (s5);
\end{tikzpicture}
\caption{An example bipartite agent interaction graph with $5$ speakers and $5$ listeners.}
\end{figure}

\end{document}